\documentclass[twocolumn,aps,showpacs]{revtex4}
\usepackage{amsmath}
\usepackage{graphicx}
\usepackage{hyperref}
\begin{document}

\title{Cohesion induced deepening transition of avalanches}

\author{Chun-Chung Chen}

\affiliation{Department of Physics, University of Washington, Seattle,
  Washington 98195}

\date{\today}

\begin{abstract}
  A directed avalanche model with a control parameter is introduced to
  describe the transition between cohesive and noncohesive granular
  material.  The underlying dynamics of the process can be mapped to
  interface growth model.  In that representation, a continuous phase
  transition separates the rough phase and the flat phase.  In the
  avalanche formulation, this corresponds to a transition from deep to
  shallow avalanches.  The scaling exponents of the avalanches indeed
  follow those of the underlying interface growth in both phases and
  at the transition point.  However, the mass hyperscaling relation is
  broken at the transition point due to the fractal nature of the
  avalanche and a hierarchy of critical directed percolation
  processes.
\end{abstract}


\maketitle

\section{Introduction}

Granular avalanches have received much attention since sandpile models
are used as paradigms of so-called self-organized criticality
\cite{Bak1987}.  However, observations of critical-type distributions
of avalanches in real physical systems are still rare, with as a
notable exception the recent rice pile experiments by Frette {\it et
  al.}~\cite{Frette1996}.  It was suggested by Christensen {\it et
  al.}~\cite{Christensen1996} that the anisotropy in the rice grains
allows more stable packing configurations in a granular pile, and that
this could be responsible for the successful observation of
criticality.  Some of the recent attention has been drawn to
avalanches in cohesive granular materials with the premise that
cohesion will also allow the sand more packing configurations and thus
increase the likelihood of observing critical scaling behavior.  While
the goal of finding criticality in cohesive sandpiles remains to be
fulfilled even after the experimental work by Quintanilla {\it et
  al.}~\cite{Quintanilla2001}, the effect of cohesion in granular
avalanches represents an interesting direction for a theoretical
study.

\begin{figure}
  \includegraphics[width=0.8\columnwidth]{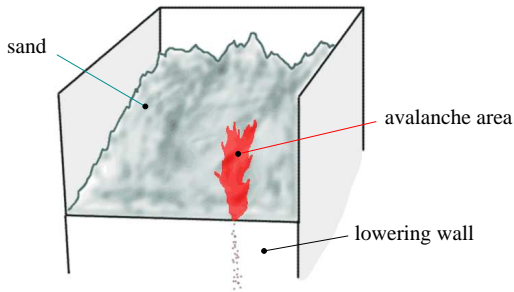}
\caption{
  A sandbox system.  The rectangular box is filled with sand.  One of
  the retaining wall can be lowered slowly to let out the sand in a
  sporadic way forming distinct avalanche events. }
\label{sandbox picture}
\end{figure}

In this article, we'll use the discrete-height version of the sandbox
(DHSB) model introduced in Ref.~\cite{Chen2002} for an unloading
sandbox (Fig.~\ref{sandbox picture}) to understand the effects of
cohesion in directed avalanche systems.
In the following section, we'll discuss how we can model cohesiveness
in avalanche systems.  In Sec.~\ref{Sec:DHSB}, we'll review the DHSB
model and introduce a cohesion parameter.  Previous results in
Refs.~\cite{Chen2002,Chen2002a} represent a special case of the model
where the system is in the deep avalanche phase with the cohesion
parameter $p=1/2$.  In Sec.~\ref{Sec:IFD}, we describe the step-flow
random-deposition (SFRD) interface growth model which underlies the
DHSB model and the directed percolation (DP) roughening transition of
the SFRD model.  In Sec.~\ref{Sec:DL}, we focus on the two
deterministic limits of the model and present the exact solution in
one of these limits.  In Sec.~\ref{Sec:FP}, numerical results for the
avalanches in the flat phase of the interface model are presented.  In
Sec.~\ref{Sec:DPRT}, we investigate the scaling behavior at the
transition point where the interface roughness increases
logarithmically in time.  We show that the avalanche-scarred sand
surface, while being rougher than nonscarred ones, retains the same
scaling exponent of the roughness in the thermodynamic limit.
However, we'll also show that at the transition point, the violation
of mass hyperscaling relation spoils the reduction to two independent
exponents established in Ref.~\cite{Chen2002a}.  We'll summarize our
results in Sec.~\ref{Sec:SM}.

\section{Tunable parameter for cohesion}

One interesting characters of cohesion in sand is that it possesses
hysteresis behavior.  Consider building a sand castle on a beach.
It's common sense that we'll need to add water to the sand before we
can shape it into a standing castle.  However, without disturbance,
the sand castle can somehow maintain its shape even after it dries out
\cite{Hornbaker1997}.  The moisture in sand increases the cohesion
between the sand particles \cite{Nase2001} and allows one to
manipulate the sand into a stable shape that, while not as attainable,
is more or less an equally valid stable shape for dry sand.

In accounting for this standing-sand-castle effect, we'll use the same
stability condition for all cohesiveness of the sandbox.  While, in
reality, the space of possible stable configurations for wet and dry
sand should not be exactly identical, in this article, we shall ignore
this distinction to avoid complicating the rules too much.

On the other hand, the way an unstable sand surface topples surely
depends on the cohesiveness.  In the DHSB model discussed below, there
are only two possible final stable states for any toppling site.
We'll call them the minimal stable state and the maximal stable state.
These two states are similar to the angle of repose and maximal stable
angle in a real sandpile.  However, in sandbox model, these states are
microscopic while the ``angles'' of a real sandpile are macroscopic.
We'll use a parameter $p$, which is a real number between $0$ and $1$,
to represent the strength of cohesion.  In the model, $p$ is the
probability for a toppling site of the sandpile to settle into the
maximal stable state instead of the minimal one.  For wet sand, the
$p$ is large, and for dry sand, the $p$ is small.

\section{Discrete-height sandbox model}\label{Sec:DHSB}

\begin{figure}
  \includegraphics[width=0.8\columnwidth]{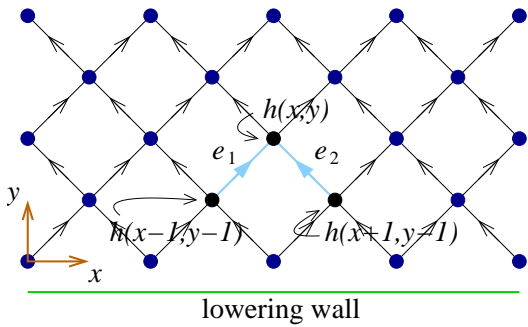}
\caption{
  The lattice structure of the two-dimensional discrete-height sandbox
  model. It corresponds to a top view of the sandbox with the lowering
  wall located at the bottom. }
\label{SB Lattice}
\end{figure}

With the discussion of the previous section in mind, let's review the
dynamic rules of the discrete-height sandbox model.  The surface of a
sandbox (see Fig.~\ref{sandbox picture}) is represented by an integer
height variable $h$ defined on a two-dimensional square lattice which
is tilted at $45^\circ$ with respect to the lowering wall as
illustrated in Fig.~\ref{SB Lattice}.
This is equivalent to considering only the lattice points whose
integer $x$ and $y$ coordinates satisfy the condition that $x + y$ is
an even number.  The lowering wall that drives the system by creating
unstable sites is located at the $y=0$ row and the activities in the
system propagate only in the positive $y$ direction.  In our numerical
simulations, the system is periodic in the $x$ direction, which is
parallel to the driving wall.  The sizes of the system in the $x$ and
$y$ directions are denoted by the numbers of sites $L_x$ in each row
and the number of rows $L_y$ respectively.

As in most sandpile processes, the dynamics of the sandbox model is
defined by a stability condition, a toppling rule, and a driving
method.  They are as follows.  The stability condition of the DHSB is
given by
\begin{equation}
 h(x,y)\leq\min\left[h(x-1,y-1),h(x+1,y-1)\right]+s_c
\label{stability5}
\end{equation}
with $s_c = 1$, which represents the local maximal stable slope.  The
unstable sites in the system topple with the rule
\begin{equation}
 h(x,y)\rightarrow\min\left[h(x-1,y-1),h(x+1,y-1)\right]+\eta,
\label{toppling5}
\end{equation}
where $\eta = 0$ with probability $1 - p$ and $\eta = 1$ with
probability $p$.  (In the earlier studies \cite{Chen2002, Chen2002a},
the value of $p$ is always $1/2$ .)  This is the only place in the
dynamics of the DHSB that the cohesion parameter $p$ comes into play.
The lowing wall which drives the system is implemented in the model by
randomly picking one of the highest sites $(x_i,0)$ on the $y=0$ row
and by reducing its height by $1$:
\begin{equation}
 h(x_i,0)\rightarrow h(x_i,0) - 1,
\label{driving5}
\end{equation}
where $i$ is the Monte Carlo time, which also serves as an age index
for the avalanches.

\begin{figure}
  \includegraphics[width=\columnwidth]{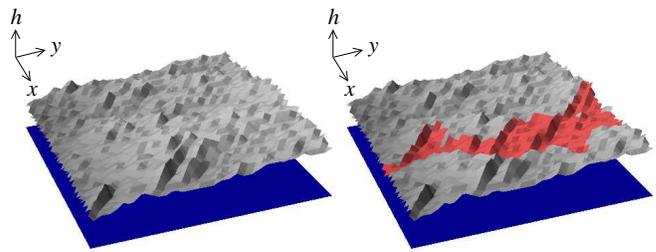}
\caption{ A typical
  configuration of the discrete-height sandbox model before (left) and
  after (right) a system spanning avalanche.  Sites participated in
  the avalanche are shaded darker.  The system size $L_x\times L_y$ is
  $32\times 64$. }
\label{Typical avalanche of DHSB}
\end{figure}

A typical configuration of the DHSB before and after an avalanche is
shown in Fig.~\ref{Typical avalanche of DHSB}.  Since the toppling of
a site on a given row $y$ only affects the stability of the two sites
immediately above it at the $y+1$ row, we choose to update the system
in a row-by-row fashion.  For each avalanche, the entire system is
stabilized by such a single sweep of topplings from $y=0$ to $y=L_y$.

\section{Underlying interface dynamics} \label{Sec:IFD}

The underlying interface dynamics of the sandbox models is given by
the step-flow random-deposition (SFRD) models with a two-step growth
rule \cite{Chen2002, Chen2002a} as illustrated in Fig.~\ref{DH SFRD
  growth}.
\begin{figure}
  \includegraphics[width=0.8\columnwidth]{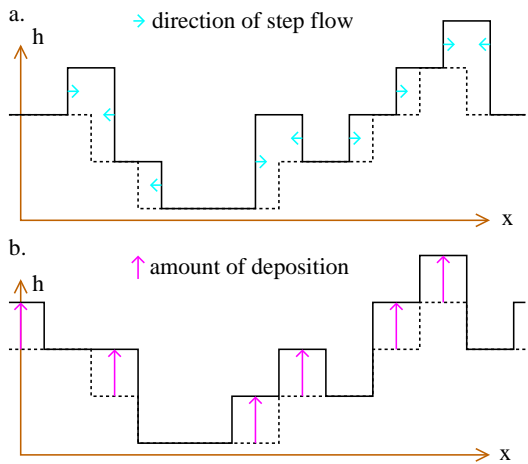}
\caption{
  Two-step growth of the discrete-height step-flow random-deposition
  interface growth model; (a) Steps flow by one unit to the right
  (left) when its size $\Delta h$ is negative (positive); (b) Each
  site increases by one unit with a probability $p$.}
\label{DH SFRD growth}
\end{figure}
The mapping between the sandbox system and the interface growth model
involves identifying the $y$ coordinate of the sandbox model with the
time $t$ of the interface growth.  Each stable sandbox surface thus
can be viewed as a space-time world-sheet configuration of the
interface growth.  Models similar to this generally belong to the
Kardar-Parisi-Zhang (KPZ) universality class \cite{Kardar1986} with
the critical exponents $\alpha = 1/2$, $\beta = 1/3$, and $z
=\alpha/\beta = 3/2$ which characterize the scaling of interface
roughness
\begin{equation}
 W^2\equiv \overline{(h-\bar h)^2}.
\end{equation}
Starting from a flat interface at $t=0$, the interface grows rougher
with
\begin{equation}
  W \sim t^\beta .
  \label{beta def}
\end{equation}
And, after a characteristic time $t_c\approx L^z$, the roughness will
saturate with a value
\begin{equation}
  W\sim L^\alpha
  \label{alpha def}
\end{equation}
depending on the system size $L$.

From the mapping introduced in Ref.~\cite{Chen2002}, the avalanche
exponents are given by
\begin{equation}
 \tau_l = \frac{\sigma - 1 -\alpha}{z} = 2,
\label{tl5}
\end{equation}
\begin{equation}
 \tau_w=\sigma - z -\alpha = \frac{5}{2},
\label{tw5}
\end{equation}
and
\begin{equation}
 \tau_\delta = \frac{\sigma - 1 - z}{\alpha} = 4
\label{td5}
\end{equation}
for the distribution functions, $P_l(l)\sim l^{\tau_l}$, $P_w(w)\sim
w^{\tau_w}$, and $P_\delta(\delta)\sim \delta^{\tau_\delta}$, of
avalanche length $l$, width $w$, and depth $\delta$.  As defined in
Ref.~\cite{Chen2002}, the avalanche length $l$ (width $w$) represents
maximum $y$ ($x$) distance of the toppling sites from the triggering
point while the avalanche depth $\delta$ is the maximum height change
of the toppling sites.  The $\sigma$ in these expressions was
eliminated with the mass hyperscaling relation
\begin{equation}
 \sigma = 2 + z + 2\alpha.
\label{hyper5}
\end{equation}
obtained from the compactness of the avalanche clusters, i.e.,
assuming $m\sim lw\delta$.

However, the discrete-height version of the SFRD model undergoes a DP
roughening transition at $p=p_c\approx 0.294515$ similar to those
studied by Kert\'esz and Wolf~\cite{Kertesz1989} also Alon {\it et
  al.}~\cite{Alon1996}.  The KPZ scaling behavior only applies when
the value of the control parameter $p$ is greater than the critical
value $p_c$.  Below this transition point the interface is in a
trivial flat state, where, for a stationary interface (interface time
$y\rightarrow\infty$), the density of sites at the bottom $h=h_0$
layer is finite.  The interface is thus \emph{pinned} at this level
and its growth rate becomes zero.

At the transition point $p=p_c$, we find the roughness of the SFRD
interface diverges only logarithmically in time
\begin{equation}
 W^2\sim (\ln t)^\gamma,
\end{equation}
with the exponent $\gamma\approx 1$ similar to that of the Kert\'esz
and Wolf's model as well as the restricted version of the models by
Alon {\it et al.}.

\section{Deterministic limits} \label{Sec:DL}

\begin{figure}
  a.\\
  \includegraphics[width=0.8\columnwidth]{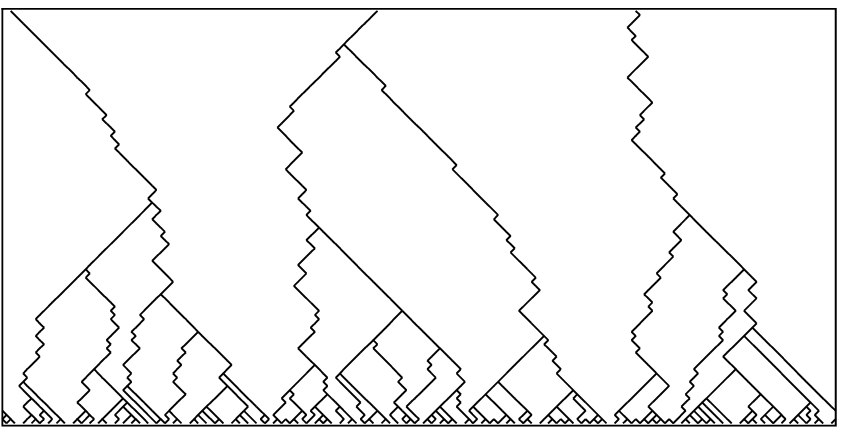}
  
  b.\\
  \includegraphics[width=0.8\columnwidth]{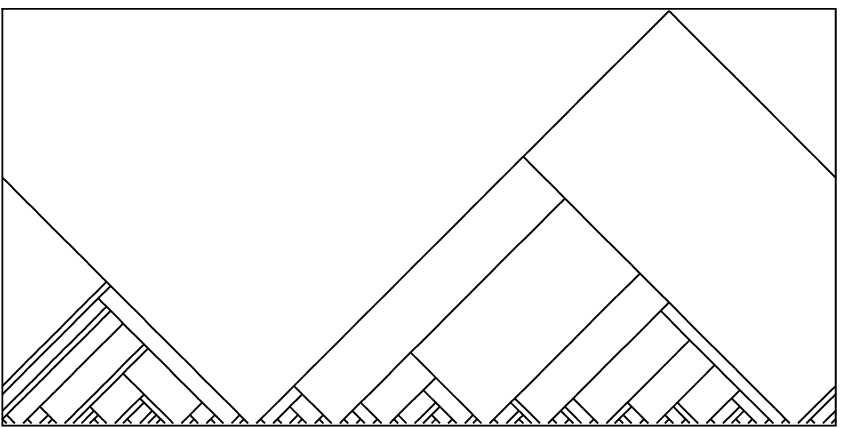}
\caption{
  Scar (edge lines of avalanche clusters) configurations of DHSB
  avalanches at the two deterministic limits; (a) $p=0$; (b) $p=1$.}
\label{deterministic scars}
\end{figure}

In the two limits, $p=1$ and $p=0$, the toppling process of the
avalanches becomes deterministic and the sand surface topples down
layer by layer.  The only randomness in the process comes from the
driving method (\ref{driving5}), i.e., that we randomly lower one of
the highest sites at the $y=0$ row to trigger an avalanche.  The
typical avalanche scar configurations at these two limits are shown in
Fig.~\ref{deterministic scars}.  These are the edges of avalanche
clusters left on the surface, some of which are partially erased by
newer avalanches.

\subsection{Domain walls at $p=0$}

\begin{figure}
  \includegraphics[width=0.8\columnwidth]{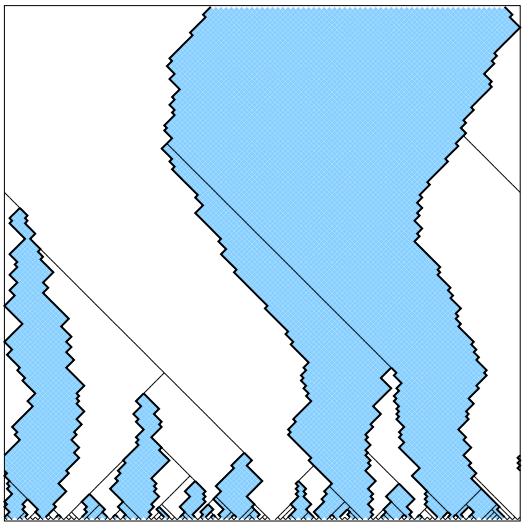}
\caption{
  The domains of odd (shaded region) and even (light region) sites on
  a DHSB surface at $p=0$, separating them are domain walls that no
  avalanche will penetrate at this deterministic limit.}
\label{domains at p0}
\end{figure}

The $p=0$ limit runs into the complication that in the bulk of the
system ($y>0$) the sand surface goes down by $2$ units at a time.
Since $\Delta h \equiv h(x,y)-\min[h(x-1,y-1),h(x+1,y-1)]= 1$ is
stable according to the stability condition (\ref{stability5}), and
the sites on the $y = 0$ row always goes down by $1$ unit each time
according to the driving method (\ref{driving5}), the sites on the $y
= 1$ row will only topple when their heights are $2$ units higher than
the triggering sites and they alway go down by $2$ units to the same
height of the triggering site according to the toppling rule
(\ref{toppling5}).  All the sites at higher rows will be locked into
the same even-oddness as the sites triggering their toppling.
Therefore, after all sites have participated in at least one
avalanche, their even-oddness will be fixed for all subsequent
topplings.  This means the even-oddness of a site is preserved by the
toppling process, and that the lines separating the even and odd sites
thus form impenetrable domain walls for the avalanches (see
Fig.~\ref{domains at p0}).  This hinders the applicability of the same
type of analysis as presented below for the $p=1$ limit.
However, the numerical results in Sec.~\ref{Sec:FP} will show that the
same scaling exponents as those of $p=1$ case control this limit, too.

\subsection{Exact solution at $p = 1$}

The $p = 1$ limit has a nice solution. Since the sites in the bulk
topple from $\Delta h = 2$ to $\Delta h = 1$, the sand surface indeed
goes down only one layer at a time without the complications as the
$p=0$ case.  An exact solution can be obtained by considering the
avalanches taking place in such one single layer.  For a brand-new
layer, the two boundaries of the first avalanche open up linearly
until the avalanche spans the system in the $x$ direction and leaves
two scar lines on the surface.  The two boundaries of the second
avalanche expand until they meet the scar lines created by the first
avalanche.  Then, they turn and follow those scar lines until they
meet with each other and terminate the avalanche.  Subsequent
avalanches follow the same scenario.  The maximum distance an
avalanche cluster can expand from its triggering point to each side in
the $x$ direction is exactly half the distance from the nearest
triggering point of the previous avalanches in the same layer on that
side.  As the triggering points are chosen in an uncorrelated manner,
the maximum width $w$ of an avalanche should follow the Poisson
distribution
\begin{equation}
 P_w(w) = \frac{\mu^w e^{-\mu}}{w!}
\end{equation}
if $\mu$ is the average distance between the triggering points of the
previous avalanches in the same layer in the stationary state.  The
avalanche under consideration could be any one of the avalanches
happening in the same layer.  Thus, we need to average over the number
of avalanches $n$ taking place before this one in the same layer.  For
a system of transverse size $L_x$, $n = L_x/\mu$, the integral can be
carried out explicitly and gives
\begin{equation}
 \int_0^\infty\frac{\mu^w e^{-\mu}}{w!}d\frac{1}{\mu} =
 \frac{(w-2)!}{w!} \sim w^{-2},
\end{equation}
which results in
\begin{equation}
 \tau_l = \tau_w = 2.
\end{equation}

The same results can also be derived from Eq.~(\ref{tl5}) and
(\ref{tw5}) by assuming $z=1$ and $\alpha=0$.  Since the avalanches
are compact, the hyperscaling relation (\ref{hyper5}) and other
exponent relations (\ref{tl5})--(\ref{td5}) from Ref.~\cite{Chen2002}
hold.

\section{Shallow-avalanche phase} \label{Sec:FP}

\begin{figure}
  \includegraphics[width=0.8\columnwidth]{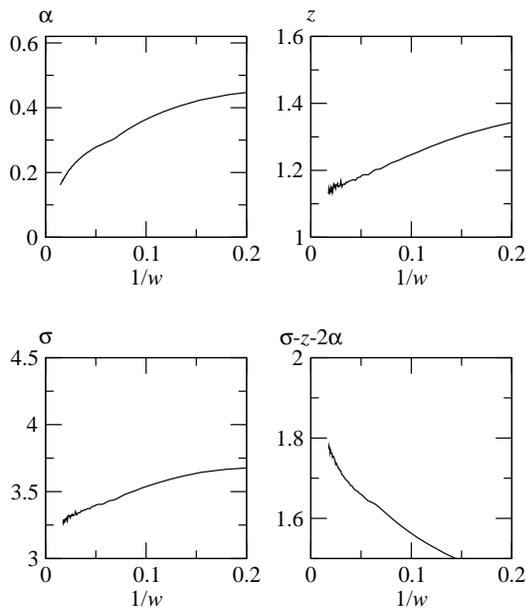}
\caption{
  Finite-size scaling (FSS) estimates of the scaling exponents versus
  inverse width ($1/w$) of avalanche clusters for the DHSB avalanches
  in the shallow-avalanche phase (measured at $p=0.1$).  They are
  consistent with $\alpha=0$ and $z=1$.}
\label{flat avalanche plots}
\end{figure}

Below the transition point, the underlying interface model is in a
flat phase where the bottom layer percolates with finite density.  All
the information of the initial configuration of the interface (the
$y=0$ row next to the wall) is wiped out at a time scale proportional
to the sizes of the islands higher than the bottom layer in the
initial state.  (Without deposition, the sizes of these islands
decrease linearly in time.)  While the underlying interface model is
in a trivial phase, much like the uncorrelated stationary state in
Dhar and Ramaswamy's directed sandpile model \cite{Dhar1989}, the
avalanche distributions of the system may still exhibit power-law
scaling.  The numerical values of the scaling exponents shown in
Fig.~\ref{flat avalanche plots} confirm the power-law scaling of the
distributions and they are similar to those values found at the $p=1$
fixed point following $z=1$ and $\alpha=0$.
While an exact solution is not available in this phase, we can
understand the scaling exponent $z=1$ from the perspective that the DP
clusters triggered from single seeds in the percolating phase open up
linearly $l\sim w$; and also that roughness exponent $\alpha=0$ comes
from that the interface is flat.  However, a difference is that while
$p<p_c$ represents an entire phase of shallow avalanches which should
be controlled by an attractive fixed point, the $p=1$ fixed point is
unstable in the sense that the scaling behavior falls back to the KPZ
universality class for any small deficiency in the cohesiveness $p$
from the value $1$.

\section{DP roughening transition} \label{Sec:DPRT}

At the transition point $p=p_c$, the interface roughness diverges
logarithmically thus the $\beta$ and $\alpha$ exponents, defined by
Eqs.~(\ref{beta def}) and (\ref{alpha def}), are both zero.
Nonetheless, the dynamic exponent $z$ has a nontrivial value
$z_\mathrm{DP}\approx 1.582$ originating from the DP nature of the
bottom-layer dynamics.  Moreover, at the transition point, the
avalanche clusters lose their compact shapes (see Fig.~\ref{typical
  avalanche cluster at DP}) and we should not expect the exponent
relations (\ref{tl5})--(\ref{hyper5}), nor the calculation in
Ref.~\cite{Chen2002a} for the corrections to scaling to remain valid.
In this section we will demonstrate the break down of mass
hyperscaling relation (\ref{hyper5}) and how the avalanches affect the
roughness of the sand surface.
\begin{figure}
  \includegraphics[width=\columnwidth]{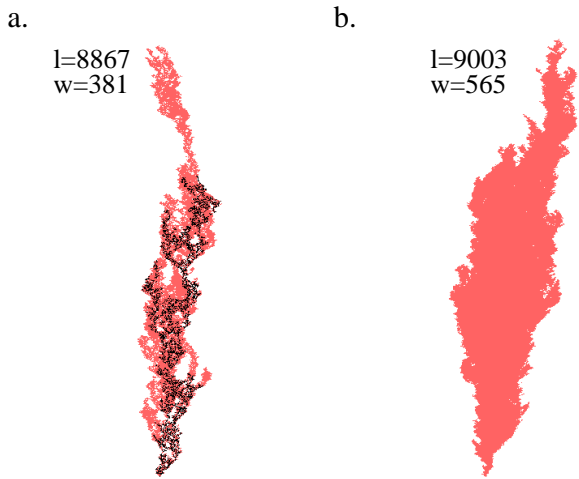}
\caption{
  Typical large avalanche cluster for DHSB (a) at the DP transition
  point; (b) in the deep avalanche phase ($p=0.5$), triggered at
  lowering wall boundary at the bottom.  The length $l$ and width $w$
  of each avalanche are as labelled.  Black area in the cluster of (a)
  is of sites that topple to the lowest height $h_0$ of the bottom
  layer.  It shows the percolation of the bottom layer.  One sees that
  the avalanche maintains a compact structure in the deep avalanche
  phase while becomes more fractal-like at the transition point.}
\label{typical avalanche cluster at DP}
\end{figure}

\subsection{Breakdown of mass hyperscaling}

At the transition point, the bottom layer of an avalanche cluster
follows the critical DP dynamics.  Therefore, we should expect from
the fractal DP cluster shape that the density of sites at the lowest
$h=h_0$ level goes to zero in the thermodynamic limit for large
avalanches.  However, the overall shape of an avalanche consists, in
addition, of sites at $h_0+1, h_0+2,$ \ldots levels.  The higher-level
sites that participate in the avalanche fill into the holes and voids
next to the bottom layer cluster and more or less bring the avalanche
cluster back to a compact shape.  We can verify this compactness of
the avalanche cluster by a direct measurement of the ratio $a/(lw)$,
with $a$ being the area of (or, the number of sites participating in)
an avalanche.  The result is shown as the solid line in Fig.~\ref{a_lw
  and m_lwd}.
\begin{figure}
  \includegraphics[width=0.8\columnwidth]{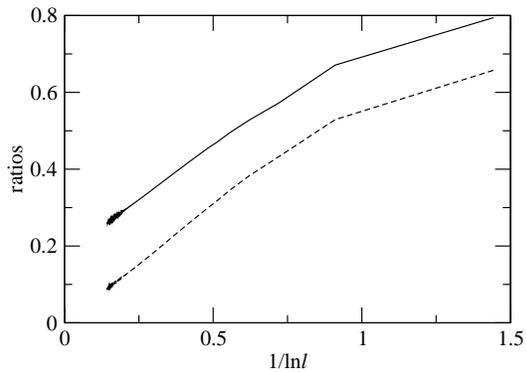}
\caption{
  The FSS plot of the area density $a/(lw)$ (solid line) and the mass
  density $m/(lw\delta)$ (dashed line) versus inverse the length
  logarithm ($1/\ln l$) for the avalanche clusters at the DP
  transition point.  While the area density converges to a finite
  value at the thermodynamic limit, the mass density converges to $0$.
}
\label{a_lw and m_lwd}
\end{figure}
The approach to a finite value on the vertical axis demonstrates the
compactness of the avalanche clusters by the existence of a finite
area density $\approx 0.2$ in the thermodynamic limit.  The FSS
estimates are plotted against $1/\ln y$ instead of $1/y$ since the
roughness of the surface diverges only logarithmically in $y$, which
will be elaborated later.

\begin{figure}
  \includegraphics[width=0.8\columnwidth]{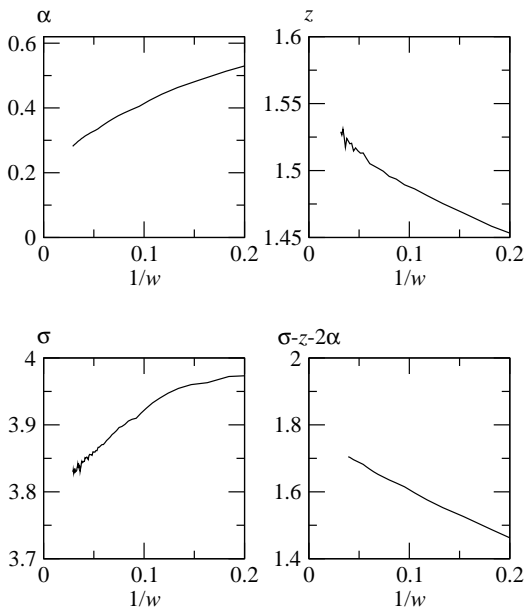}
\caption{
  FSS estimates of the scaling exponents derived from the avalanche
  exponents $\tau_l$, $\tau_w$, $\tau_\delta$ for the discrete-height
  sandbox model versus the inverse width ($1/w$) at the DP transition
  point.  The $z$ exponent is consistent with dynamic exponent of DP
  universality class $z_\mathrm{DP}\simeq 1.582$.  The combination
  $\sigma - z - 2\alpha < 2$ indicates a violation of mass
  hyperscaling relation (\ref{hyper5}).}
\label{azs2 plots DP}
\end{figure}

Contrary to a finite area density, as also shown in Fig.~\ref{a_lw and
  m_lwd}, the mass density $m/(lw\delta)$ (the dashed line) goes to
zero in the thermodynamic limit.  The absence of a finite mass density
breaks the scaling
\begin{equation}
 m\sim lwd,
\label{compact5}
\end{equation}
which lead, in Ref.~\cite{Chen2002}, to the mass hyperscaling relation
(\ref{hyper5}).  The plot of the combined exponent $\sigma - z -
2\alpha$ in Fig.~\ref{azs2 plots DP} shows the violation of
Eq.~(\ref{hyper5}) as the FSS estimates approach $\approx 1.72$ which
is much lower than the expected value $2$ for compact avalanches
obeying Eq.~(\ref{compact5}).

Also shown in Fig.~\ref{azs2 plots DP} are the plots for the $\alpha$,
$z$, and $\sigma$ exponents.  They are consistent with $z =
z_\mathrm{DP}$ and more or less with $\alpha = 0$.  This confirms that
the scaling behavior of the avalanches follows those of the SFRD
interface.  The slow convergence of $\alpha$ is to be expected from
the logarithmic divergence of the interface roughness.

\subsection{Interface roughness}

The remaining question is how the scaling behavior of the roughness is
changed by the iterated avalanche process.  We approach this by
looking at the change of the global surface roughness itself and by
comparing the scaling of this change to the scaling of the original
interface roughness.  The same analysis was performed in
Ref.~\cite{Chen2002a} which concerns only the $p=1/2$ case of the
DHSB, and it was found that the change in the global roughness by the
avalanche process only represents large corrections to the KPZ scaling
behavior of the surface.  However, at the DP transition point, the
interface roughness diverges only logarithmically.  This makes the
scaling of interface roughness more likely to be overwhelmed by the
change in the roughness due to the avalanche process, and we generally
would not expect the values of the scaling exponents to remain the
same.  In the following, we'll show the scaling of the interface does
follow the same logarithmic divergence.

We perform a direct measurement of the global interface roughness at
the transition point.  The results are shown in Fig.~\ref{roughness
  at DP}(a).
\begin{figure}
  a.\\
  \includegraphics[width=0.8\columnwidth]{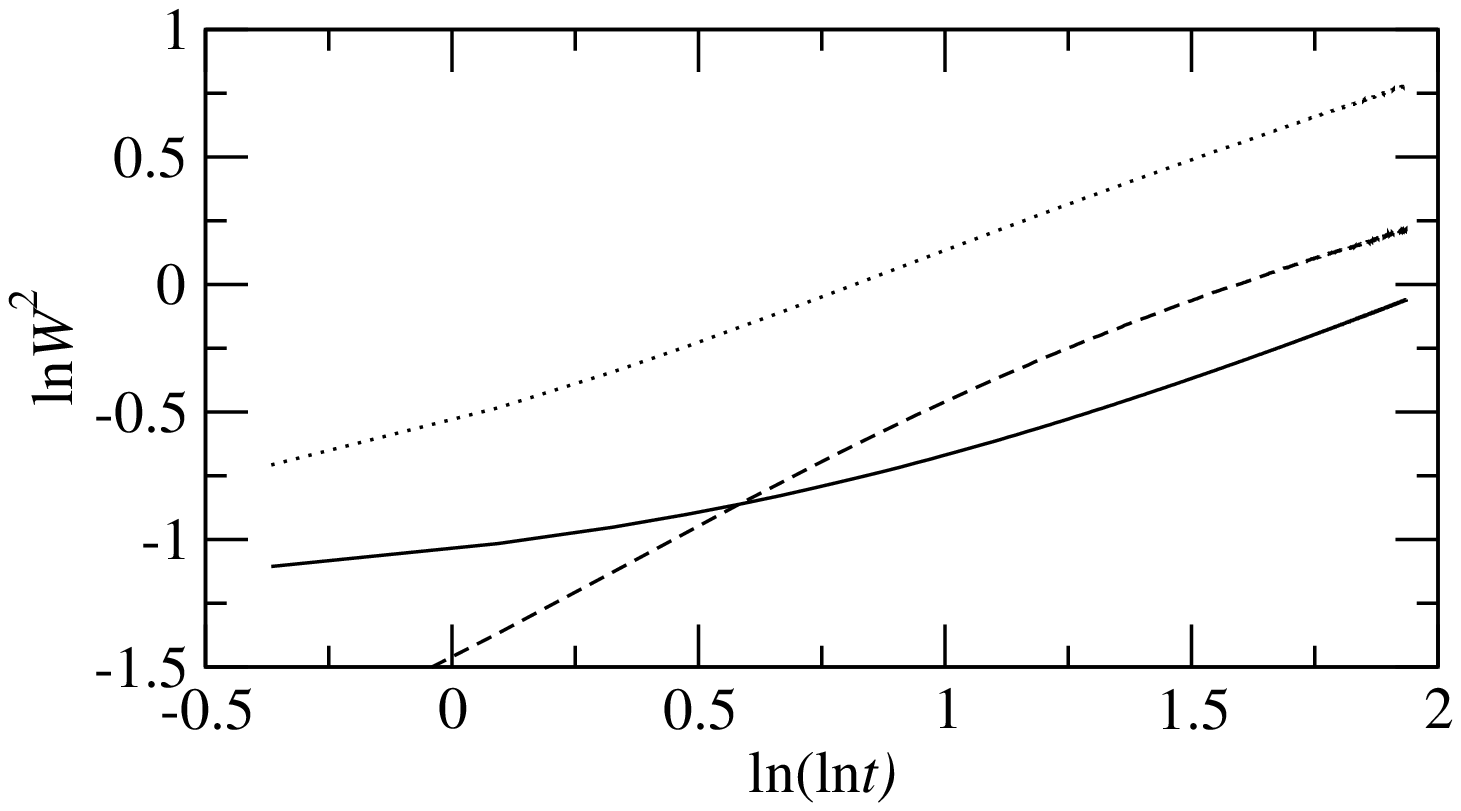}
  
  b.\\
  \includegraphics[width=0.8\columnwidth]{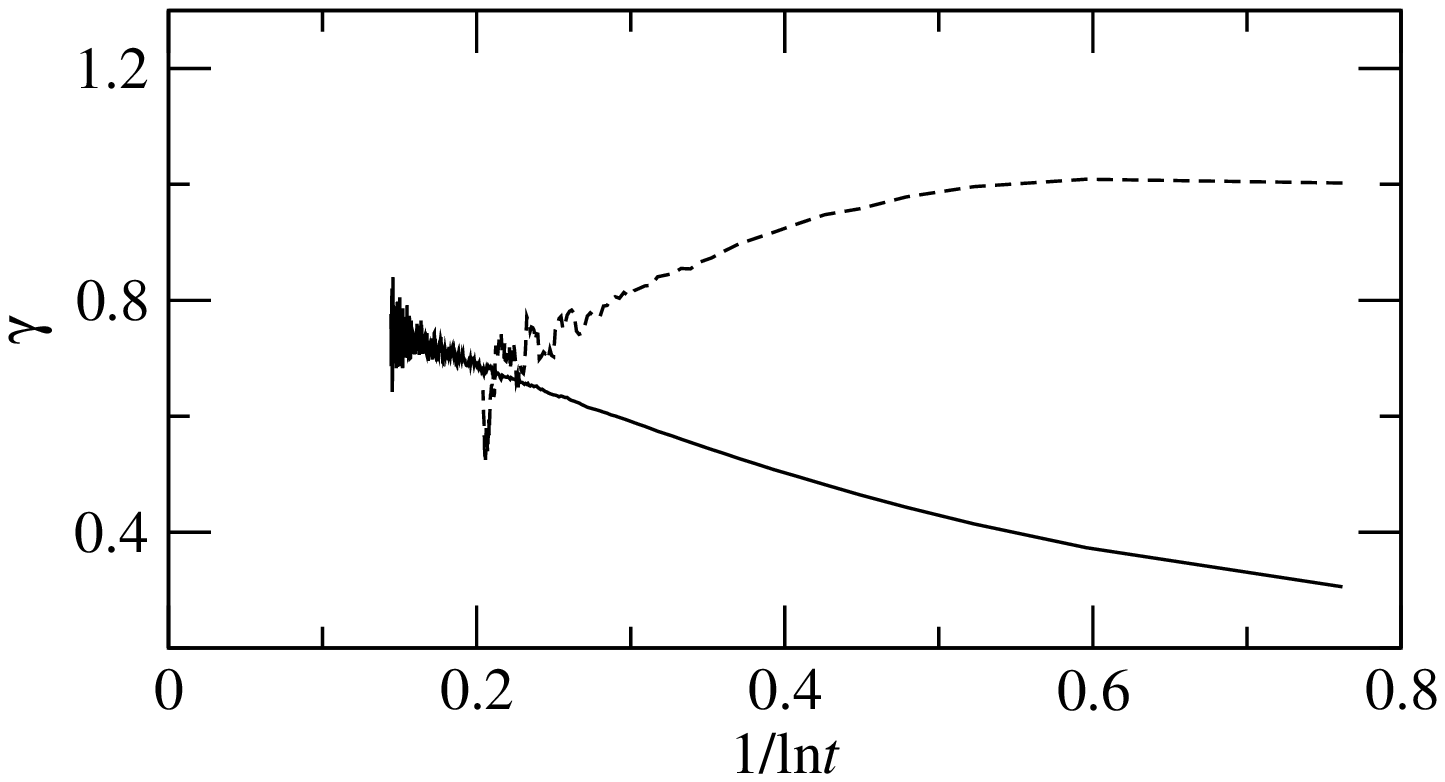}
\caption{
  (a) The roughness of a stationary DHSB surface(dotted line) compared
  with the roughness of the SFRD model (solid line) versus the double
  logarithm of time $t$ at the DP transition point.  The iterated
  avalanche process makes the surface rougher.  The dashed line shows
  the difference $\Delta W^2\equiv W_\mathrm{DHSB}^2 -
  W_\mathrm{SFRD}^2$ between the roughness of the two.  (b) FSS of the
  $\gamma$ exponents of the logarithmic scaling for the SFRD roughness
  $W_\mathrm{SFRD}^2$ (solid line) and the difference $\Delta W^2$
  (dashed line), both assumed to have the scaling form $(\ln
  t)^\gamma$, versus the inverse of the logarithm of time.  In the
  $t\rightarrow \infty$ limit, $\Delta W^2$ scales with a smaller
  $\gamma$ exponent than that of $W^2$. }
\label{roughness at DP}
\end{figure}
As in the $p=1/2$ case, the surface is made rougher by the iterated
avalanches.  The increase in the roughness $\Delta W^2$ scale as $(\ln
t)^{\gamma_\Delta}$ with the exponent $\gamma_\Delta \approx 0.4$
which is shown as the dashed line in Fig.~\ref{roughness at DP}(b).
Since the interface roughness itself scales as $W^2\sim (\ln
t)^\gamma$ with $\gamma\approx 1$ which is shown as the solid line in
Fig.~\ref{roughness at DP}(b), the change in the roughness is
irrelevant comparing to the interface scaling.  We can thus conclude
that in the thermodynamics limit, the stationary surfaces of DHSB have
the same $\gamma$ exponent as the SFRD interfaces.  Just as the in the
deep phase ($p=1/2$) of the the avalanche, the iterated avalanche
process only gives rise to sizable corrections to the interface
scaling behavior.

\section{Summary} \label{Sec:SM}

In this article, we introduced the DHSB as a model for avalanches in
granular materials with variable cohesiveness.  This model exhibits a
deepening transition from a shallow-avalanche phase where avalanches
only involve a couple of surface layers of the granular material, into
a deep-avalanche phase where the depths of avalanches increase as
power laws in their lengths or widths.  In the deep-avalanche phase,
the scaling behavior of the avalanches belongs to the KPZ universality
class: The avalanche clusters scale anisotropically with $l\sim
w^{3/2}$ and depth increase as $\delta\sim w^{1/2}$.  In the flat
phase, the avalanche clusters scale isotropically $l\sim w$ with
finite depths.

In both phases, the mass hyperscaling relation (\ref{hyper5}) based on
compactness (\ref{compact5}) of the avalanches holds.  On the other
hand, at the transition point, the hierarchical DP structure, pointed
out by T\"auber {\it et al.}~\cite{Tauber1998}, for each height level
breaks this scaling in a subtle way.  While the mass density
$m/(lw\delta)$ of the avalanche clusters goes to zero in the
thermodynamic limit, the area density $a/(lw)$ remains finite.
However, the exact scaling behavior of the systems at this DP
roughening transition point remains unclear even without the iterated
avalanche in the DHSB model \cite{Alon1998, Lopez1998,
  Goldschmidt1999}.

While we are not aware of any experimental study on how the avalanche
behavior of a system will vary with a gradual change in the
cohesiveness of the grains, the cohesiveness in granular system is
known to vary with moisture \cite{Nase2001} and grain sizes
\cite{Valverde2000, Quintanilla2001}.  We thus expect experimental
studies in this direction to be feasible.  The DHSB model represents a
system with a layered structure where the heights are discrete, and
the DP nature of the deepening transition relies heavily on a
well-defined bottom layer or minimal stable configuration of the
system.  It thus wouldn't be a surprise if exact DP scaling were not
to be observed in the avalanches of most experimental sandpiles.
Nonetheless, the breakdown of the mass hyperscaling relation
(\ref{hyper5}) comes from the fractal aspect of the hierarchical DP
clusters and is a more fundamental property.  It would serve as a
hallmark of such a transition if it's to be observed experimentally.

The author thanks Marcel den Nijs for many helpful discussions and
critical reading of the manuscript.  This research is supported by the
National Science Foundation under Grant No. DMR-9985806.


\begin{thebibliography}{10}

\bibitem{Bak1987}
P. Bak, C. Tang, and K. Wiesenfeld, Phys. Rev. Lett. {\bf 59},  381  (1987).

\bibitem{Frette1996}
V. Frette {\it et~al.}, Nature {\bf 379},  49  (1996).

\bibitem{Christensen1996}
K. Christensen {\it et~al.}, Phys. Rev. Lett. {\bf 77},  107  (1996).

\bibitem{Quintanilla2001}
M.~A.~S. Quintanilla, J.~M. Valverde, A. Castellanos, and R.~E. Viturro, Phys.
  Rev. Lett. {\bf 87},  194301  (2001).

\bibitem{Chen2002}
C.-C. Chen and M. {den Nijs}, Phys. Rev. E {\bf 65},  031309  (2002).

\bibitem{Chen2002a}
C.-C. Chen and M. {den Nijs}, Phys. Rev. E {\bf 66},  011306  (2002).

\bibitem{Hornbaker1997}
D.~J. Hornbaker {\it et~al.}, Nature {\bf 387},  765  (1997).

\bibitem{Nase2001}
S.~T. Nase, W.~L. Vargas, A.~A. Abatan, and J.~J. McCarthy, Powder Technology
  {\bf 116},  214  (2001).

\bibitem{Kardar1986}
M. Kardar, G. Parisi, and Y.-C. Zhang, Phys. Rev. Lett. {\bf 56},  889  (1986).

\bibitem{Kertesz1989}
J. Kert\'esz and D.~E. Wolf, Phys. Rev. Lett. {\bf 62},  2571  (1989).

\bibitem{Alon1996}
U. Alon, M.~R. Evans, H. Hinrichsen, and D. Mukamel, Phys. Rev. Lett. {\bf 76},
   2746  (1996).

\bibitem{Dhar1989}
D. Dhar and R. Ramaswamy, Phys. Rev. Lett. {\bf 63},  1659  (1989).

\bibitem{Tauber1998}
U.~C. T\"auber, M.~J. Howard, and H. Hinrichsen, Phys. Rev. Lett. {\bf 80},
  2165  (1998).

\bibitem{Alon1998}
U. Alon, M.~R. Evans, H. Hinrichsen, and D. Mukamel, Phys. Rev. E {\bf 57},
  4997  (1998).

\bibitem{Lopez1998}
J.~M. L\'opez and H.~J. Jensen, Phys. Rev. Lett. {\bf 81},  1734  (1998).

\bibitem{Goldschmidt1999}
Y.~Y. Goldschmidt, H. Hinrichsen, M. Howard, and U.~C. T\"auber, Phys. Rev. E
  {\bf 59},  6381  (1999).

\bibitem{Valverde2000}
J.~M. Valverde, A. Castellanos, A. Ramos, and P.~K. Watson, Phys. Rev. E {\bf
  62},  6851  (2000).

\end{thebibliography}
\end{document}